\documentclass[a4paper, amsfonts, amssymb, amsmath, reprint, showkeys, nofootinbib, twoside, longbibliography]{revtex4-2}
\usepackage[english]{babel}
\usepackage[utf8]{inputenc}
\usepackage[colorinlistoftodos, color=green!40, prependcaption]{todonotes}

\usepackage{natbib}

\usepackage{amsthm}
\usepackage{mathtools}
\usepackage{physics}
\usepackage{xcolor}
\usepackage{graphicx}
\usepackage{bbm}
\usepackage[paperheight=11in, paperwidth=8.5in,left=23mm,right=13mm,top=35mm,columnsep=15pt]{geometry}
\usepackage{adjustbox}
\usepackage{placeins}
\usepackage[T1]{fontenc}
\usepackage{lipsum}
\usepackage{csquotes}
\usepackage{mathrsfs}

\newcommand{\diag}[1]{\mathrm{diag}{#1}}

\newcommand\restr[2]{{
  \left.\kern-\nulldelimiterspace 
  #1 
  \vphantom{\big|} 
  \right|_{#2} 
  }}

\usepackage[pdftex, pdftitle={Article},pdfauthor={Author}]{hyperref} 
\usepackage{fullpage}
\usepackage{bbold}
\usepackage{indentfirst}
\usepackage{braket}
\usepackage{physics}
\usepackage{graphicx}
\usepackage{commath}
\usepackage{tikz-cd}
\usepackage{hyperref}

\usepackage{caption}
\usepackage{subcaption}
\usepackage{ragged2e}

\makeatletter
\def\@fnsymbol#1{\ensuremath{\ifcase#1\or\dagger\or \ddagger\or
   \mathsection\or \mathparagraph\or \|\or **\or \dagger\dagger
   \or \ddagger\ddagger \else\@ctrerr\fi}}
\makeatother

\setcitestyle{super}

\makeatletter
\newcommand*{\citens}[2][]{%
  \begingroup
  \let\NAT@mbox=\mbox
  \let\@cite\NAT@citenum
  \let\NAT@space\NAT@spacechar
  \let\NAT@super@kern\relax
  \renewcommand\NAT@open{[}%
  \renewcommand\NAT@close{]}%
  \cite[#1]{#2}%
  \endgroup
}
\makeatother

\begin{document}
\captionsetup[figure]{name={FIG.},labelsep=period,justification=raggedright}
\title{The Optimization of Flux Trajectories for the Adiabatic Controlled-Z Gate on Split-Tunable Transmons}

\author{Vihaan Dheer\\\textit{vdheer@students.hackleyschool.org}}

\address{Hackley School. 293 Benedict Avenue, Tarrytown, NY 10591, USA.}


\begin{abstract}
In a system of two tunable-frequency qubits, it is well-known that adiabatic tuning into strong coupling-interaction regions between the qubit subspace and the rest of the Hilbert space can be used to generate an effective controlled-Z rotation. We address the problem of determining a preferable adiabatic trajectory along which to tune the qubit frequency, and apply this to the flux-tunable transmon model. The especially minimally anharmonic nature of these quantum processors makes them good candidates for qubit control using non-computational states, as long as higher-level leakage is properly addressed. While the statement of this method has occurred multiple times in literature, there has been little discussion of which trajectories may be used. We present a generalized method for optimizing parameterized families of possible flux trajectories and provide examples of use on five test families of one and two parameters.
\end{abstract}


\maketitle

\section{Introduction} \label{sec:outline}
With the rapid development of quantum computation and information theory, it is not uncommon that implementation issues arise from approximations in theoretical considerations\cite{nielson, preskill2021quantum, sep-qm-decoherence}. There is an ensemble of negative effects causing quantum states to decohere, including state relaxation, leakage, dephasing, and other environmental coupling factors which are often difficult to control\cite{nielson, computing_decoherence}. It is thus crucial to minimize these errors both at their source, and, if possible, after they have acted on a system. In the latter case, quantum error correction\cite{quantum_error_correction} provides information-theoretic means for resolving these issues in certain cases, though physical correction is usually desirable. One source of error hard to mitigate is that of leakage out of the computational subspace, which is especially difficult to address on transmon superconducting qubits\cite{superconducting_qubits}. In the so-called transmon regime, anharmonicity is kept relatively small to maintain low sensitivity to charge noise in the superconducting circuit. This, however, considerably increases the probability of unwanted transitions out of the computational subspace. Much of the field's literature treats these non-computational states as pure error; here we discuss a case in which their consideration is used to provide an advantage.

It is well-known\cite{Strauch_2003, DiCarlo_2009, Martinis_2014} that the use of the avoided crossing between the $\ket{1,1}$ and $\ket{2,0}$ states of coupled transmons in an adiabatic fashion leads to a CZ operation, and in general any controlled phase (throughout this work, we use the definition $CZ=\mathcal{I}-2\ketbra{1,1}$). The subject of this paper, motivated by the potential \textit{use} of non-computational states in tunable transmons, is an in-depth model addressing the optimization of flux trajectories for this application. Most work on the subject explains only the possibility of this implementation, but does not specify which trajectories to use. An explicit analysis of adiabatic control waveforms was carried out by Martinis and Geller\cite{Martinis_2014}, however this was done with a focus towards the \textit{generation} of such trajectories and keeping short gate times. Though short operations  are indeed important to building a successful quantum gate, here we will fix a gate time, focusing more on the reduction of leakage, which, in addition to increasing the fidelity of the gate, also improves performance of future operations in a gate sequence\cite{superconducting_qubits}.

Specifically, in a mathematically rigorous manner, after devising a functional norm on the algebra of trajectories to determine to what degree a trajectory is adiabatic \textit{without} any direct quantum mechanical simulation, we analyze multiple trajectories which should theoretically implement a CZ gate. We introduce a generalized method to locate optimal trajectories in families of curves.

The structure of this paper is as follows: we first discuss our model of the two-transmon quantum processor and the theoretical underpinning for the adiabatic implementation of the controlled-Z gate in Sec. \ref{sec:model} and Sec. \ref{sec:error}. In Sec. \ref{sec:main_analysis} and \ref{sec:par_opt}, respectively, we present our mathematical model and subsequently define test trajectories. Lastly, we compare the action of two of these through simulation of two tunable transmons in Sec. \ref{sec:simulation} and conclude by discussing possible future work in Sec. \ref{sec:conclusion}.

\section{Model \& Theory} \label{sec:model}
\begin{figure}[t]
    \centering
    \includegraphics[width=\linewidth]{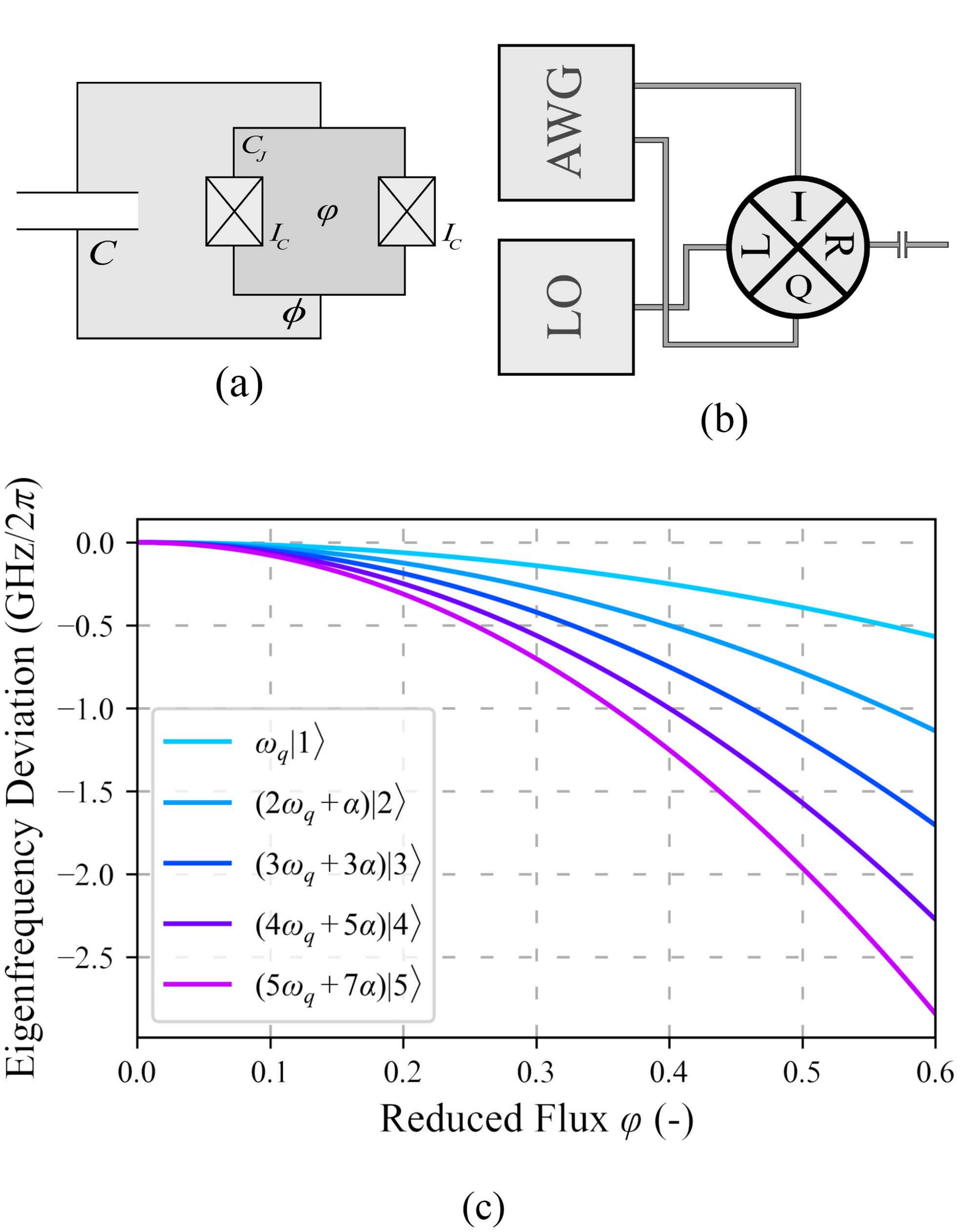}
    \caption{\textbf{(a)} A simplified schematic of a single flux-tunable transmon circuit with shunt capacitance $C$ and junction self-capacitance $C_J$. An external reduced flux $\varphi$ is threaded through a dc-SQUID (i.e. a split-junction) to vary the Josephson energy\cite{superconducting_qubits} $E_J(\varphi) = 2I_C\abs{\cos{\varphi}}2\pi/\Phi_0$. \textbf{(b)} A simplified microwave drive line setup, similar to those described in Refs. \citenum{superconducting_qubits, z_gates}. An AWG shapes pulses generated by a local oscillator on in-phase $I$ and quadrature $Q$ components, sending outputs through an IQ mixer to a line capacitively coupled with a transmon. $\textbf{(c)}$ The effect of threading a flux through the split-junction of a transmon with anharmonicity $\alpha$; the higher energy levels of the transmon experience more frequency drop from the tuning.}
    \label{fig:tunable_transmon}
\end{figure}

In this section, we briefly describe the model that the rest of this work is based on, which is used for both simulation and motivation of our optimization method. In addition, we review how a controlled phase gate naturally arises in the adiabatic control of the system when levels in the qutrit subspace of the transmon are considered. In our setup, two flux-tunable transmons are employed in the standard arrangement: the qubit frequency is tuned by applying magnetic flux through the symmetric split-junction (a dc-SQUID)[Fig. \ref{fig:tunable_transmon}(a)]. We use simulation parameters similar to those in Ref. \citenum{sim_params}, in which Q1(Q2) has frequency $\omega/2\pi=5.889 (5.031)$ GHz and anharmonicity $\alpha/2\pi=-324.3 (-234.7)$ MHz, with coherence times $T1=25.5 (48.8)$ $\mu$s and $T2=13.3 (28.4)$ $\mu$s, with coupling strength between qubits $g/2\pi=24.7$ MHz. Though a common choice at present for a coupling setup is cQED\cite{superconducting_qubits}, for analytical simplicity we use direct capacitive coupling between the qubits, yet the strategies presented here are generalizeable to a cQED model (such as the generalized Tavis-Cummings Hamiltonian\cite{DiCarlo_2009}). In this configuration, single-qubit microwave pulses are implemented through the capacitive coupling of each transmon to a drive line controlled by an arbitrary waveform generator[Fig. \ref{fig:tunable_transmon}(b)]\cite{z_gates}. The operational point for these microwave pulses occurs when there is no applied flux, i.e. $\varphi=0$; at this base point, the qubit frequencies of the transmons are furthest detuned from each other, thus minimizing the effect of the coupling as a source of decoherence. Fig. \ref{fig:tunable_transmon}(c) shows the explicit effect of the flux detuning on the frequency for various initial states. In the tensor product eigenbases of the individual transmons (including coupling to their own drive lines), the Hamiltonian governing the capacitive interaction is\cite{superconducting_qubits}
\begin{equation}
    H=\sum_j{\Big(\omega^j_1\ketbra{j}{j}\otimes\mathcal{I}_2 + \omega^j_2\mathcal{I}_1\otimes \ketbra{j}{j}\Big) - g(a - a^\dagger)^{\otimes 2}\label{two_qubit_interaction}}
\end{equation}
where $g$ is the coupling strength between the transmons, $\omega^j_1$ and $\omega^j_2$ are the respective qubit frequencies of $j$th level of each transmon, $\mathcal{I}_i$ is the identity operator on the Hilbert space for the $i$th transmon, and we have used the usual oscillator annihilation and creation operators $a$ and $a^\dagger$. Since the coupling $g/2\pi$ is significantly smaller (roughly 200-fold at $\varphi=0$) than the qubit frequencies, the rotating wave approximation is justified, reducing the interaction term of Eq. \ref{two_qubit_interaction} to
\begin{equation}
    H_\mathrm{int} = g(a\otimes a^\dagger + a^\dagger\otimes a). \label{h_int_eq}
\end{equation}
A controlled-Z gate being the application of a $\pi$ phase to $\ket{1,1}\equiv\ket{1}\otimes\ket{1}$, any of its physical realizations must be able to address $\ket{1,1}$ differently than $\ket{01}$ and $\ket{10}$. In the qubit subspace $\mathcal{H}_{qq}\subset\mathcal{H}$ (where $\mathcal{H}=\mathrm{dom}{(H)}$ is the coupled-transmon Hilbert space which the Hamiltonian in Eq. \ref{two_qubit_interaction} operates on), the eigenfrequencies corresponding to $\ket{0,1}$ and $\ket{1,0}$ simply sum to that of $\ket{1,1}$, and any CZ gate would need to be realized through external considerations. If we do \textit{not} restrict ourselves to $\mathcal{H}_{qq}$, however, then the minimally anharmonic nature of transmons and the capacitive coupling differentiates the state $\ket{1,1}$ from the noted sum, and the possibility for addressing it separately opens up. Unfortunately, as discussed, the system rests in the single-qubit operational point, where the Hamiltonian can be approximated by $H_1\otimes\mathcal{I}_2+\mathcal{I}_1\otimes H_2$; thus the CZ operational point, at flux $\varphi_Z>0$, must be one where qubit frequencies are close enough to allow for relatively strong coupling interactions. 

In what follows, by $\omega_{ij}$ we mean the the eigenvalues of $H$ \textit{corresponding} to the $\ket{i,j}$ state, for which at $\varphi=0$, $\omega_{00}=0,\omega_{01}\approx\omega^2_1,\omega_{10}\approx\omega^1_1$, and as stated, $\omega_{11}\approx\omega^1_1+\omega^2_1$. In addition, when discussing the matrix representation of an operator on a subspace of $\mathcal{H}$, we will work in a basis ordered by eigenvalues, which does not preserve standard qubit ordering when its domain is restricted to $\mathcal{H}_{qq}$. To naturally create a CZ operation, we shall work in a seven-dimensional subspace of $\mathcal{H}$, namely $\mathcal{H}_Z\subset\mathcal{H}$, spanned by the basis set $\mathcal{B}_7=\{\ket{i, j}|i,j\leq 2\}-\{\ket{0, 0}, \ket{2,2}\}$. We exclude $\ket{0, 0}$ and $\ket{2,2}$ from this consideration since $H_\mathrm{int}$ as defined in Eq. \ref{h_int_eq} when restricted to the full qutrit subspace does not have any matrix elements that generate interaction in the first or ninth row or column (explicitly, $\bra{i,j}\!{H_\mathrm{int}}\!\ket{2,2}=\bra{2,2}\!{H_\mathrm{int}}\!\ket{i,j}=\bra{i,j}\!{H_\mathrm{int}}\!\ket{0,0}=\bra{0,0}\!{H_\mathrm{int}}\!\ket{i,j}=0\;\forall i,j\leq 2$ holds in the restriction to the qutrit subspace). For both notational simplicity and possible interpretational advantage, we write the Hamiltonian by means of block matrices, utilizing the well-known Pauli matrices and the lesser-known Gell-Mann matrices, whose use in a qutrit-like consideration is somewhat fundamental since, as the Pauli matrices span the Lie algebra $\mathfrak{su}(2)$, the eight Gell-Mann matrices span $\mathfrak{su}(3)$. For a complete list, description, and interpretation of these matrices, see Ref. \citenum{gell-mann_matrices}; here we will only use them for convenience of notation (with the symbol $\lambda$). Eq. \ref{h_int_eq} under this restriction (and the earlier mentioned eigenvalue ordering) becomes
\begin{equation}
    \restr{H_\mathrm{int}}{\mathcal{H}_Z}\equiv H_\mathrm{int}\simeq\begin{pmatrix}
    g\sigma_1   &   0   &   0 \\
    0 & g\sqrt{2}(\lambda_1+\lambda_6) & 0 \\
    0 & 0 & 2g\sigma_1
    \end{pmatrix}.\label{matrix_h_int}
\end{equation}
on $\mathcal{B}_7$ (for operator-matrix equivalence in some specified basis we use the symbol $\simeq$). This operator representation hints at contributions important to mitigate in the execution of CZ: for example, the $g\sigma_1$ term represents a swapping action between $\ket{1,0}$ and $\ket{0,1}$ excitations at the coupling strength. In fact, this term can be used to naturally induce an iSWAP gate\cite{iSWAP_impl}, however this should be avoided for a successful and coherent CZ. In addition, the $g\sqrt{2}\lambda_1$ and $g\sqrt{2}\lambda_6$ terms represent the main issue, i.e. swapping between $\ket{0, 2}$ and $\ket{1,1}$, and $\ket{2, 0}$ and $\ket{1,1}$ respectively; in fact only the latter will be an issue, since we shall drive the transmon close to a point at which swapping is highly likely. Lastly, we have another swapping interaction with strength $2g$ between $\ket{1,2}$ and $\ket{2,1}$, however ideally these should not matter much if leakage out of $\mathcal{H}_{qq}$ has been well-managed throughout the quantum gate sequence.
\begin{figure}[t]
    \centering
    \includegraphics[width=\linewidth]{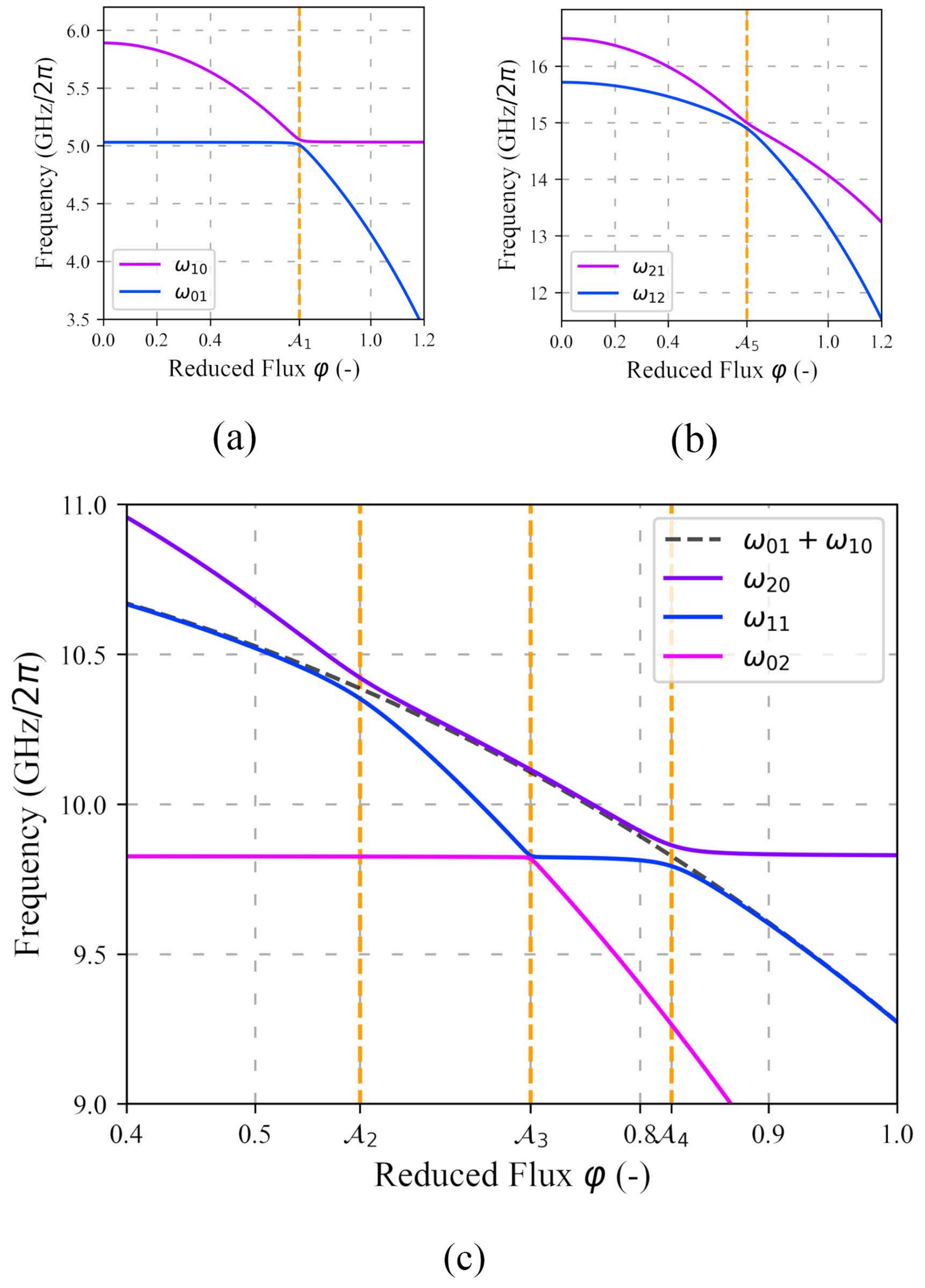}
    \caption{Different ranges of the spectral decomposition of the Hamiltonian in Eq. \ref{two_qubit_interaction} as applied flux is varied. \textbf{(a)} Shows an avoided crossing $\mathcal{A}_1$ between $\ket{0,1}$ and $\ket{1,0}$; tuning the system to $\mathcal{A}_1$ naturally generates an iSWAP operation (see main text). \textbf{(b)} A higher-level avoided crossing $\mathcal{A}_5$ between $\ket{1,2}$ and $\ket{2,1}$. \textbf{(c)} Shows three avoided crossings, the important one being $\mathcal{A}_2$ between $\ket{2,0}$ and $\ket{1,1}$. For comparison, we have in order, $\mathcal{A}_2=0.5815988,\mathcal{A}_5=0.6951558,\mathcal{A}_3=0.7145798,\mathcal{A}_1=0.7334230,\mathcal{A}_4=0.8241256$.}
    \label{fig:spectral_decomp_one}
\end{figure}

Fig. \ref{fig:spectral_decomp_one} shows the low, medium, and high frequency ranges of the spectral decomposition of $H_\mathrm{int}$, in which (a) and (b) show two avoided crossings useful for inducing various state changes. Most importantly, (c) gives an idea of how the CZ is implemented: the black dashed line, the sum of $\omega_{01}$ and $\omega_{10}$, deviates from $\omega_{11}$ as flux is increased, which occurs because of the downward push on $\ket{1,1}$ from $\ket{2,0}$. The optimal flux bias for CZ is $\mathcal{A}_2$, the avoided crossing between these two states, or just before it\cite{Martinis_2014}, since being in the $\ket{1,1}$ state and moving towards $\mathcal{A}_2$ maximizes both accumulation of the difference in its frequency and that of the previously noted sum \textit{and} the probability of remaining in $\ket{1,1}$ during the movement. As we have assumed coupling is minimal at zero flux bias, we can approximate the eigenstates of $H$ at $\varphi=0$ to be the individual transmon eigenstates. Parameterizing $H$ by $\varphi$, if $\partial_\varphi H\approx 0$, i.e. the trajectory $\varphi(t)$ is adiabatic, then the system stays in its instantaneous eigenstate\cite{adiabatic_theorem} at $\varphi(0)=0$. The rotating frame of the transmon at which the remainder of the quantum gate sequence is operated in is defined at the zero flux-bias point, so the Hamiltonian \textit{does} have time-dependent components, but only along its diagonal. Since the Hamiltonian commutes with itself at different times, the Schrodinger equation and the adiabatic approximation dictates that the trajectory-dependent unitary $U_A:[0,\infty)^T\to\mathcal{B}(\mathcal{H}_Z)$
\begin{equation}
    U_A(\varphi)\simeq
    \begin{pmatrix}
    e^{i\Theta_{01}(\varphi)} & 0 & \cdots &  0 \\
    0 & e^{i\Theta_{10}(\varphi)} & \cdots & 0  \\
    \vdots & \vdots & \ddots & \vdots           \\
    0 & 0 & \cdots & e^{i\Theta_{21}(\varphi)}
    \end{pmatrix}
\end{equation}
(here $\mathcal{B}(\mathcal{H})$ denotes the Banach space of bounded linear operators on a Hilbert space $\mathcal{H}$) is applied to $\mathcal{H}_Z$ in the basis $\mathcal{B}_7$, where $T=\mathrm{dom}(\varphi)$ is the time interval over which $\varphi$ acts, and
\begin{equation}
    \Theta_{ij}(\varphi)=\int_T{\!\delta_{ij}(\varphi(t))\,dt},
\end{equation}
where for some $t$, $\delta_{ij}(\varphi(t))=\omega_{ij}(0)-\omega_{ij}(\varphi(t))$ is the flux-dependent deviation of the eigenfrequency corresponding to $\ket{i,j}$. We conclude this section by noting that, if we choose some flux trajectory $\varphi_\pi$ such that $\Theta_{11}(\varphi_\pi)-\Theta_{01}(\varphi_\pi)-\Theta_{10}(\varphi_\pi)\equiv\pi\pmod{2\pi}$, then the CZ operation is realized up to global phase as the product of the restriction of $U_A$ and single-qubit gates, namely
\begin{equation}
    \restr{R_3(\Theta_{10}(\varphi_\pi))\otimes R_3(\Theta_{01}(\varphi_\pi))}{\mathcal{H}_A}\restr{U_A(\varphi_\pi)}{\mathcal{H}_A}\sim_\phi \restr{CZ}{\mathcal{H}_A},\label{equiv_to_cz}
\end{equation}
where $\mathcal{H}_A=\mathcal{H}_Z\,\cap\,\mathcal{H}_{qq}$, the equivalence relation $\sim_\phi$ is defined for operators $A,B\in\mathcal{B}(\mathcal{H})$ on any Hilbert space $\mathcal{H}$ by $A\sim_\phi B$ iff $A=e^{i\theta}B$ for some $\theta\in[0,2\pi)$, and we have used standard qubit rotation operators $R_j(\theta)=\exp(-i\pi\theta\sigma_j/2)$. Note that although Eq. \ref{equiv_to_cz} is written for $\mathcal{H}_A$, the three-dimensional subspace of $\mathcal{H}$ involving all two qubit states except for $\ket{0,0}$, it would also hold for the full $\mathcal{H}_{qq}\supset\mathcal{H}_A$ if the $\ket{0,0}$ state is included in $\mathcal{B}_7$, yet it is not relevant to $U_A$, which leaves $\ket{0,0}$ unchanged under this extension.

\section{Error in Adiabatic Approximation} \label{sec:error}
We next seek to mathematically quantify the leakage out of the computational subspace after the adiabatic transition; this manifests itself as error in the adiabatic approximation, which we derive first in the general case and then specify it to the leakages most important in this adiabatic passage. Given some time-dependent Hamiltonian $H(t)$ acting on a Hilbert space $\mathcal{H}$, we let $\{\ket{\phi_n(t)}\colon n\in\mathbbm{Z}^+\}$ be the set of satisfiers of the instantaneous eigenstate condition $H(t)\ket{\phi(t)}=E(t)\ket{\phi(t)}$. Since these form a complete orthonormal basis of $\mathcal{H}$, we can write a general solution to the Schrodinger equation as
\begin{equation}
    \ket{\psi(t)}=\sum_n{c_n(t)\ket{\phi_n(t)}}
\end{equation}
for constants $c_n(t)\in\mathbbm{C}$. Applying the Schrodinger equation and taking the inner product with $\bra{\phi_k(t)}$, rearranging gives\cite{adiabatic_theorem}
\begin{equation}
    i\hbar\frac{\mathrm{d}c_k}{\mathrm{d}t}=\left(E_k(t)-i\hbar\braket{\phi_k(t)\vert\dot{\phi}_k(t)}\right)c_k(t)-i\hbar\mathcal{E}_{k}(t),
\end{equation}
where $\mathcal{E}_{k}$ represents the error in the adiabatic approximation i.e. leakage for instantaneous eigenstate $\ket{\phi_k}$, defined by
\begin{equation}
    \mathcal{E}_{k}(t)=\sum_{n\neq k}{c_n(t)\braket{\phi_k(t)\vert\dot{\phi}_n(t)}}.\label{error_term}
\end{equation}
This follows intuition as if $\mathcal{E}_k(t)\approx 0$, a system that starts in some $\ket{\phi_k(0)}=\ket{\psi(0)}$ will remain in the same eigenstate. Thus, to minimize leakage out of the computational subspace, we wish to minimize the inner product in Eq. \ref{error_term}. In the case we are interested in, after differentiating the instantaneous eigenstate condition and manipulating, the error can be rewritten as
\begin{equation}
    \mathcal{E}_{k}(t)=\sum_{n\neq k}{c_n(t)\dot{\varphi}(t)\frac{\bra{\phi_k(t)}\partial_\varphi H\ket{\phi_n(t)}}{\hbar\omega_n(\varphi(t))-\hbar\omega_k(\varphi(t))}}.
\end{equation}
In summary, minimizing the leakage involves, as expected, minimizing the time derivative of the flux trajectory, yet it is also important to keep energies of states at risk of coupling far from each other, as well as it is to minimize the corresponding matrix element of $\partial_\varphi H$. 

\section{Theoretical Trajectory Design} \label{sec:main_analysis}
The issue with the idealized CZ gate discussed in Sec. \ref{sec:model} above lies in the adiabatic approximation: perfect adiabaticity is not possible since $\partial_t H$ cannot non-trivially be the zero operator. In this section we shall discuss how to quantify adiabaticity for a trajectory $\varphi(t)$ using arguments related to the form of the Hamiltonian in Eq. \ref{matrix_h_int}. We shall work under the following assumptions and assertions for $\varphi\colon T\to\mathbbm{R}^+$:
\begin{enumerate}
    \item $\varphi\eval_{\partial T}=0$. Inaction at the endpoints ensures that the qubit frequency starts and ends up in the optimal single-qubit operational point where coupling between transmons is negligible.
    \item $\int_T{\!\zeta(\varphi(t))\,dt}\equiv\pi\pmod{2\pi}$, where $\zeta(\varphi)=\delta_{11}(\varphi)-\delta_{10}(\varphi)-\delta_{01}(\varphi)$. As mentioned in the previous section, this condition generates the relative phase applied to $\ket{1,1}$ during tuning.\label{req:zeta}
    \item The set of implementable flux pulses is exactly $C_0(T)$. That is, any $\varphi(t)$ is physically continuous, and any $C_0$ function may be flux-implemented.
    \item $\varphi(t)$ does not induce a $\ket{1,1}\longleftrightarrow\ket{2,0}$ transition with high likelihood, i.e. $\bra{2,0(t)}\partial_\varphi H\ket{1,1(t)}\approx0$.\label{req_no_tran}
    \item $\varphi(t)$ does not induce a $\ket{0,1}\longleftrightarrow\ket{1,0}$ transition with high likelihood, i.e. $\bra{1,0(t)}\partial_\varphi H\ket{0,1(t)}\approx0$.
\end{enumerate}
Incidentally, the last requirement is satisfied by keeping $\varphi(t)<\mathcal{A}_1\:\forall t\in T$ as shown in Fig. \ref{fig:spectral_decomp_one}(a), and we can at least partially satisfy condition \ref{req_no_tran} by setting $\varphi(t)\leq \mathcal{A}_2$, thus we restrict the codomain of $\varphi$ to $[0,\mathcal{A}_2]$. Keeping these conditions in mind, we look to the $\mathcal{H}_Z$ Hamiltonian in Eq. \ref{matrix_h_int}, and transform into the rotating frame set by $H(\varphi=0)$, which has `central' matrix element
\begin{multline}
\cos{(\Delta t-\alpha_2 t)}\lambda_1+\sin{(\Delta t - \alpha_2 t)}\lambda_2
+ \cos{(\Delta t+\alpha_1 t)}\lambda_6 \\+ \sin{(\Delta t + \alpha_1 t)}\lambda_7
-\diag{(0,\delta,2\delta)}, \label{long_exp}
\end{multline}
where $\alpha_1$ and $\alpha_2$ are the anharmonicities of the transmons, $\delta$ is the deviation of the qubit frequency of transmon one, and $\Delta$ is the detuning between the qubit frequencies of the two transmons at $\varphi=0$. The eigenfrequency difference $\zeta(\varphi)$ is most efficiently raised when $\varphi$ is kept near the $\mathcal{A}_2$ avoided crossing as in Fig. \ref{fig:spectral_decomp_one}(c), and most time should be spent there in the ideal case. This point is unfortunately the most dangerous with respect to condition \ref{req_no_tran}, and we seek to develop a rough ansatz for how to construct $\varphi(t)$ such that this is carefully handled. We see that the unwanted transitions occur in the latter terms of Eq. \ref{long_exp}, which, when restricted to the interaction subspace $\mathcal{H}_I$ spanned by $\{\ket{1, 1},\ket{2, 0}\}$, produces the propagator $U_\mathrm{Ev}(t)\in\mathcal{U}(\mathcal{H}_I)$, where
\begin{equation}
    U_{\mathrm{Ev}}(t) = \mathcal{T}\exp{-i\int_0^t{\!\begin{pmatrix}
-\delta(t') & e^{-i(\Delta+\alpha_1)t'} \\
e^{i(\Delta+\alpha_1)t'} & -2\delta(t')
\end{pmatrix}\,dt'}}
\end{equation}
where $\mathcal{T}$ is the Dyson time ordering operator\cite{sakurai} (and we use the notation $\mathcal{U}(V)$ to signify the group of unitary operators on a vector space $V$). Ideally the time-evolution restricted to $\mathcal{H}_I$ is just the two-dimensional identity, which should not create any swapping interaction; in this case, clearly $\delta(\varphi)$ must be minimized to remove $\ket{2,0}$ amplification, and while the off-diagonal terms cannot be easily controlled, repeated integration by parts shows that their contribution is lowered with sufficiently distant initial qubit frequencies.

We address the issue of minimizing $\delta(\varphi)$ and $\partial_t H$ simultaneously by promoting the set of continuously differentiable choices for $\varphi$, that is $S=C_1\,\cap\, [0,\mathcal{A}_2]^T$, to a semi-normed associative algebra $\mathcal{N}=(S, \|\!\cdot\!\|_{\mathcal{N}}\colon\mathcal{N}\to\mathbbm{R},+,\cdot)$ under usual function addition and multiplication. We make a simple but non-trivial choice for $\|\cdot\|_{\mathcal{N}}$, where, for constants $\upsilon, \kappa$,
\begin{equation}
    \|f\|_{\mathcal{N}}=\int_T{\!\Big\{\upsilon\delta(f(t))+\kappa\dot{\delta}(f(t))^2\Big\}\,dt}\;\;\forall f\in\mathcal{N},\label{func_norm}
\end{equation}
(note that the semi-norm may take on negative values) which defines a `large' element as one which is generates a large change in frequency and its time-derivative, i.e. one that is diabatic. Of course, this is just one arbitrary way of defining diabaticity, but it provides a good starting point for finding sufficiently adiabatic trajectories. Note that the desired trajectory $\varphi$ must additionally satisfy condition \ref{req:zeta}, and if one imagines the norm on $\mathcal{N}$ as a single-valued uncountably-infinite dimensional curve, then we are looking for the minimum of the intersection of this curve and the restriction on $\zeta(\varphi)$. This is easily approached using the Calculus of Variations, in which constrained variation via Lagrange multipliers is performed (for a rigorous treatment of this procedure see Ref. \citenum{calc_of_var}). In the general case, for a Lagrange multiplier $\lambda$, we minimize the functional $F[\varphi]$ while satisfying constraint functional $G[\varphi]=c_0\in\mathbbm{R}$ by imposing $\delta (F+\lambda G)=0$ ($\delta$ being the first variation). In our case, $F[\varphi]=\|\varphi\|_\mathcal{N}$ and $G[\varphi]=\int_T{\!\zeta(\varphi(t))\,dt}\equiv\pi\pmod{2\pi}$. If the functional sum is treated as an effective classical action, then the effective Lagrangian is
\begin{equation}
    \mathcal{L}_\mathrm{eff}(\varphi, \dot{\varphi}, t)=\upsilon\sqrt{\cos{\varphi}}-\frac{\kappa}{4}\dot{\varphi}^2\tan{\varphi}\sin{\varphi}+\lambda\zeta(\varphi),
\end{equation}
where we have used the definition for $\delta(t)$ and removed the arbitrary multiplicative factor $\omega_1-\alpha_1$ by absorbing into the Lagrange multiplier. Applying the variation gives
\begin{multline}
    \lambda\cos{\varphi(t)}\cot{\varphi(t)}\frac{\partial\zeta}{\partial\varphi}-2\frac{\upsilon}{\kappa}\sqrt{\cos^3\!\varphi(t)}\,+ \\
    (1+\cos^2\!\varphi(t))\left(\frac{d\varphi(t)}{d t}\right)^2-\sin{(2\varphi(t))}\frac{d^2 \varphi}{dt^2}=0,\label{hard_ode}
\end{multline}
a highly non-linear second order equation after absorbing more constants into $\lambda$. With this, we impose that $\varphi\eval_{\partial T}=\dot{\varphi}\eval_{\partial T}=0$ along with the constraint equation for the multiplier. This, in fact, is problematic enough to discourage physical realization of the optimal solution; the main issue is the trigonometric singularity on $\partial T$ induced by the boundary condition in the first term. We shall instead use Eq. \ref{hard_ode} as another method of gauging how diabatic a trajectory $\varphi$ is; \textit{this} equation, however, gives us a distance at each $t$ from an optimal solution curve, something more valuable than the norm presented above. As for the singularity, we note that only relative diabaticity is relevant to this discussion, and we shall thus consider trajectories that start at some small $\varepsilon\gtrsim 0$. 

\section{Parametric Optimization} \label{sec:par_opt}
\begin{figure}[t]
    \centering
    \includegraphics[width=\linewidth]{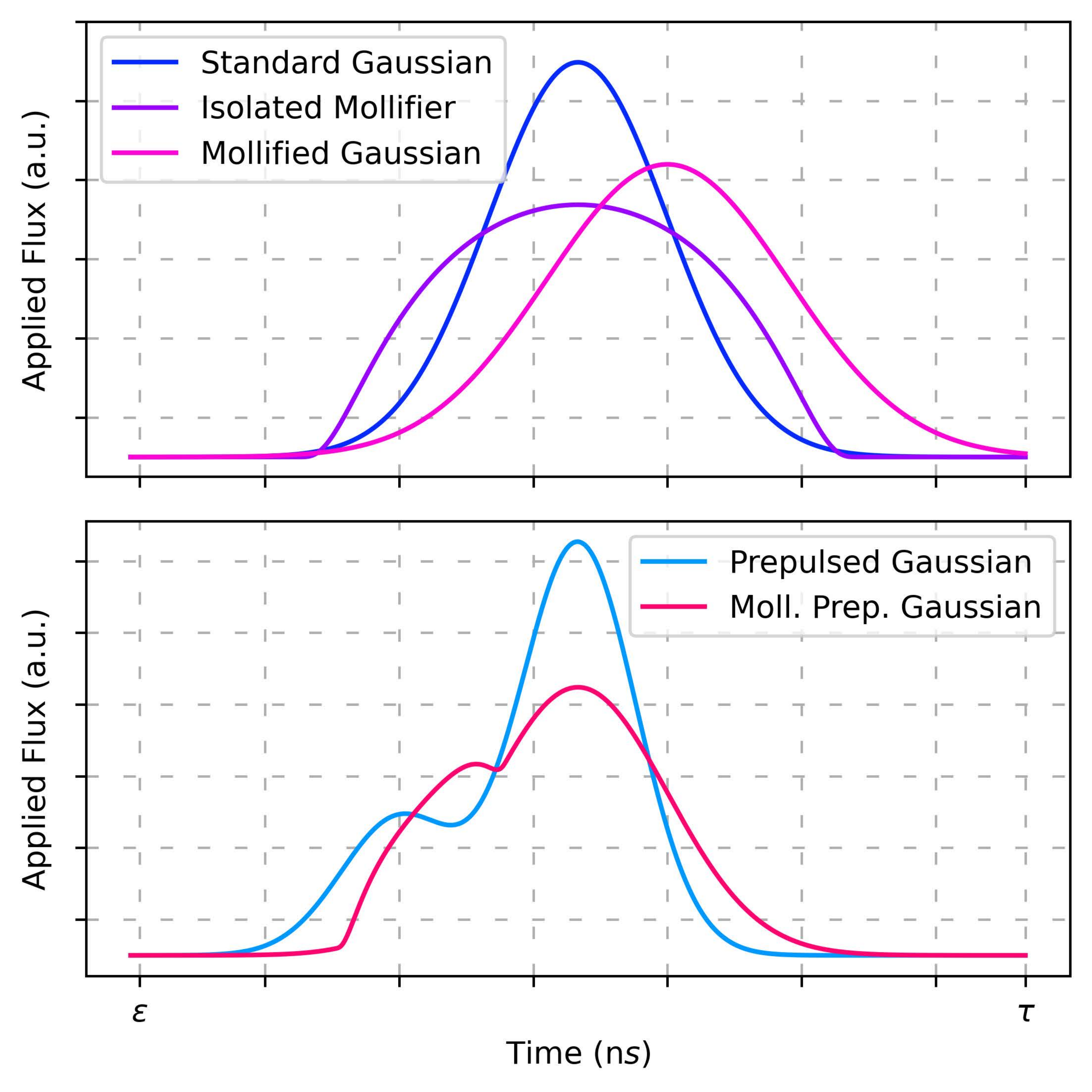}
    \caption{Flux trajectories to be tested. The parameters used to generate the example plots are consolidated in the order of the trajectory definition. For trajectories 1-5, $\sigma=2,6.5,2.5,1.3,2$ and for 3-5, $\mu=4,12,13$.}
    \label{fig:flux_traj_list}
\end{figure}

To proceed, we utilize Eq. \ref{hard_ode} through a functional operator $\mathcal{D}\colon S'\to C_1$ equal to the LHS, where $S'=C_1\,\cap\,[0,\mathcal{A}_2]^{T'}$, $T'=[\varepsilon, \tau]$, and $\tau$ is the CZ gate time. As potential flux trajectories, we shall consider a few possible curves detailed below for demonstration (all temporal units are in n$s$):
\begin{enumerate}
    \item \textit{Standard Gaussian}. For this we use the standard equation
    \begin{equation}
        \mathscr{G}(t,\sigma, \tau/2)=\mathcal{P}_{\sigma,\tau}C_{\sigma,\tau}e^{-\frac{1}{2}\left(\frac{t-\tau/2}{\sigma}\right)^2}-\mathcal{Z}_{\sigma, \tau},
    \end{equation}
    where $C_{\sigma,\tau}$ normalizes the usual part of the distribution to one, $\mathcal{P}_{\sigma,\mu}$ satisfies the $\pi\pmod{2\pi}$ requirement on $\zeta$, and $\mathcal{Z}_{\sigma,\tau}$ ensures that $\mathscr{G}(0,\sigma)=\mathscr{G}(\tau,\sigma)=0$. This will act as a control relative to the other pulses, as we expect to find that by increasing $\sigma$ (and subsequently $\tau$, otherwise there is a diabatic jump on the boundary) we can arbitrarily reduce the diabaticity, and that being the only parameter makes this a somewhat trivial case. From now on, we shall suppress the constants' dependence on $\tau$.
    \item \textit{Isolated Mollifier}. We introduce a mollifier curve, namely
    \begin{equation}
        \mathscr{M}(t,\sigma, \tau/2)=\left\{\begin{matrix}
        \mathcal{P}_\sigma C_\sigma e^{\left(\left(\frac{t-\tau/2}{\sigma}\right)^2-1\right)^{-1}} & \text{if } t\in\mathcal{W} \\
        0 & \text{otherwise}
        \end{matrix}\right.,
    \end{equation}
    an often-used object when a point of nondifferentiability must be smoothed out (here $\mathcal{W}=[\tau/2-\sigma,\tau/2+\sigma]$). Like the Gaussian waveform, this has only one width-parameter, and therefore acts again as a trivial control with respect to optimization. This curve has the advantage that all of its derivatives are exactly zero on $\partial T$.
    \item{\textit{Mollified Gaussian}} Convolving the two previous trajectories, we have a less trivial case, in which it is of interest where each curve is placed. This takes the following form:
    \begin{equation}
        \mathscr{F}(t,\sigma, \tau/2, \mu)=\int_T{\!\mathscr{G}(t, \sigma, \tau/2)\mathscr{M}(t'-t, \sigma, \mu)\, dt'}
    \end{equation}
    Adjustment of the parameter $\mu$ changes where the convolution occurs; we have also suppressed any normalization constants which maintain the $\zeta$ requirement.
    \item \textit{Prepulsed Gaussian}. We `prepulse' the above Gaussian waveform, giving
    \begin{equation}
        \mathscr{H}(t,\sigma,\tau/2,\mu)=\frac{1}{3}\mathscr{G}(t,\sigma,\mu)+\mathscr{G}(t,\sigma,\tau/2)
    \end{equation}
    by using another Gaussian at some point during the trajectory; this could relieve some of the diabaticity generated by the original curve since it concentrates some of the `mass' to a different area. Though the arbitrary factor of 1/3 was chosen as a relative height between the prepulse and the main Gaussian, the important effects should at least somewhat be observed in this trial.
    \item \textit{Mollifier-Prepulsed Gaussian} Similar to the previous curve, we again prepulse a Gaussian trajectory, however this time we do so using an isolated mollifier, which becomes
    \begin{equation}
        \mathscr{J}(t,\sigma,\tau/2,\mu)=\frac{1}{6}\mathscr{M}(t,\sigma,\mu)+\mathscr{G}(t,\sigma,\tau/2).
    \end{equation}
\end{enumerate}
These trajectories are shown in Fig. \ref{fig:flux_traj_list} with fixed example parameters.

As the purpose of these is a demonstration of the model detailed in this work, we will proceed with a visual representation of the non-simulative optimization process. In practice, when flux trajectories are further constrained by lab conditions, families of flux trajectories should be computationally optimized to determine suitable parameter values. Here, however, graphical comparison describes the procedure well enough. 

\begin{figure*}[t]
    \centering
    \includegraphics[width=\linewidth]{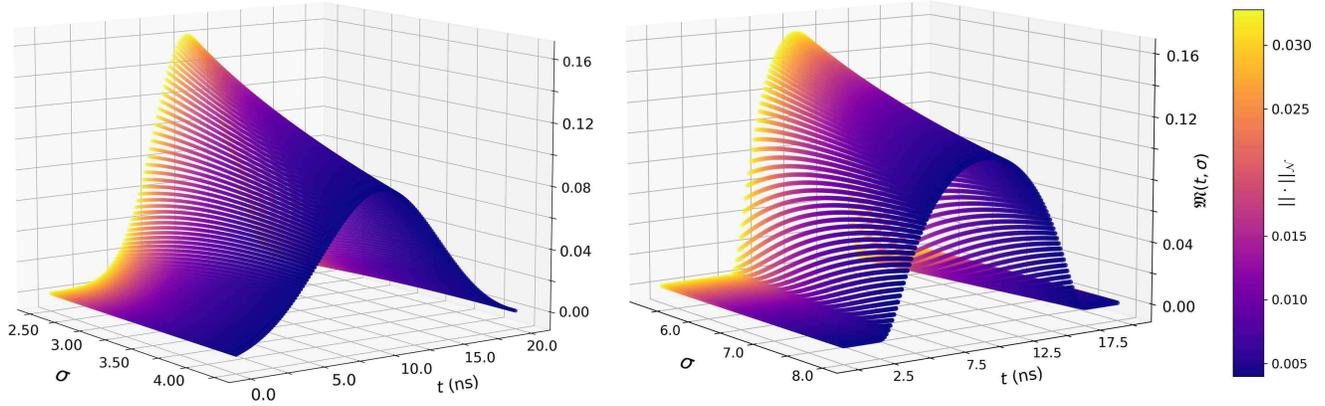}
    \caption{Standard Gaussian and Isolated Mollifier trajectories plotted as a function of $t, \sigma$ colored by $\|\cdot\|_{\mathcal{N}}$ for values of $\sigma$. Wider trajectories in each family are clearly much smaller in $\mathcal{N}$, matching intuition about adiabaticity. The $\mathcal{N}$-norm values should be considered in arbitrary units as trajectories here are normalized to one, which may not satisfy the condition on $\zeta(\varphi)$.}
    \label{single-param-traj}
\end{figure*}

\begin{figure}[t!]
    \centering
    \includegraphics[width=0.95\linewidth]{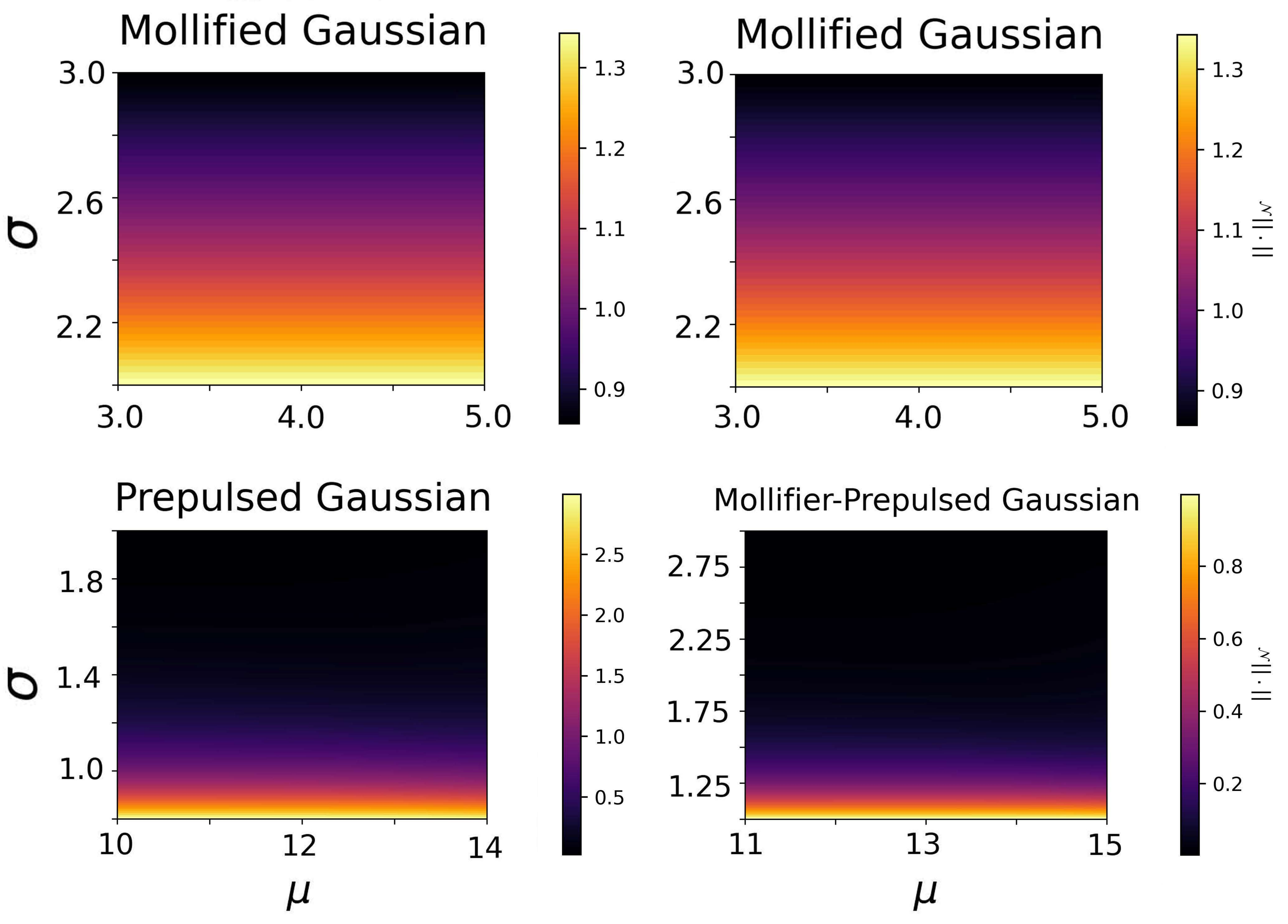}
    \caption{The value of $\mathcal{D}$ for a Standard Gaussian trajectory for $\sigma=3.75$ [top left], and plots of the $\mathcal{N}$-norm for all three doubly-parametric trajectories mentioned.}
    \label{norm-on-trajs}
\end{figure}

We can confirm our control hypothesis on the first two, singly-parametric trajectories, shown in Fig. \ref{single-param-traj}, since the $\mathcal{N}$-norm can clearly arbitrarily be reduced by choosing large values for $\sigma$. Of course, the drawback is having greater derivatives around $\partial T$; we can refer to $\mathcal{D}$ for more specificity on this, which we have computed only for the Standard Gaussian trajectory for demonstration of principle [Fig. \ref{norm-on-trajs}]. As one can see, the additional measure $\mathcal{D}$ provides insight as to at which \textit{location} a trajectory is diabatic. This not only acts as a secondary metric but also assists in adjusting families for further optimization. More physically, $\mathcal{D}$ can also aid in determining adjustments to flux trajectories which aim to decrease errors due to leakage, in accordance with the adiabatic error term $\mathcal{E}(t)$ defined in Sec. \ref{sec:error}.

For the other curves, we evaluate the $\mathcal{N}$-norm on non-temporal parameter space, shown in Fig. \ref{norm-on-trajs}. Again in accordance with expectations, we see that while shifting (varying $\mu$) does not affect the norm, we should be able to increase adiabaticity by enlarging $\sigma$. In addition, we eliminate from a consideration of optimization the mollified Gaussian trajectory due its large $\mathcal{N}$-norm in comparison to the other two trajectories.

\section{Quantum Mechanical Simulation} \label{sec:simulation}
To conclude a demonstration of this process, we can directly simulate the effects of the some of the optimal trajectories discussed in the previous section. For simplicity, we do not consider the effects of stochastic decoherence on the system, which greatly computationally eases both the simulation process and comparison of results. We make this choice almost without loss of generality, since any trends present in this analysis are likely to persist with the inclusion of decoherence into the model.

While challenging to classify the results of a simulation under a specific trajectory, we can simplify the procedure by assuming perfect linearity on the restriction to $\mathcal{H}_Z$; we thus have a representation of the operator by temporally evolving the elements of the tensor product basis set. We can construct such a matrix representation using
\begin{equation}
    \mathcal{M}(\varphi)=\sum_{\ket{e},\,\ket{f}\,\in\,\mathcal{B}_8}{\left(\bra{f}U_A(\varphi)\ket{e}\right)\ket{e}\otimes\bra{f}},
\end{equation}
where $\mathcal{B}_8=\mathcal{B}_7\,\cup\,\{\ket{0, 0}\}$. We have included the basis vector $\ket{0,0}$ to draw an accurate comparison between the ideal CPHASE and $\mathcal{M}(\varphi)$ (note that $U_A$ as used here has been extended to act on a Hilbert space, $\mathcal{H}_8$, larger than $\mathcal{H}_Z$ which is spanned by $\mathcal{B}_7$). Defining the accuracy of $\mathcal{M}$ is equivalent to choosing a matrix norm, thus for simplicity we make use of the $\ell_2$ spectral norm\cite{matrix_norm} to compute the distance from $\mathcal{M}$ to a perfect $CZ$. We define $\mathcal{F}:\mathcal{H}_8\to\mathbbm{R}^+$ such that
\begin{equation}
    \mathcal{F}(\mathcal{M}(\varphi))=\max_{\abs{\mathbf{x}}_{{\ell}_2}\neq\, 0}{\frac{\abs{\left(\mathcal{M}(\varphi)-CZ\right)\mathbf{x}}_{\ell_2}}{\abs{\mathbf{x}}_{\ell_2}}},
\end{equation}
which is computed with the standard singular value method (note that $CZ$ as used here has been extended with zeroes to eight dimensions). For the Gaussian trajectory $\mathfrak{G}$ with $\sigma=3.75$, we obtain the following simulation result (abbreviated to $\mathcal{H}_A$),
\begin{equation*}
\mathcal{M}(\mathfrak{G})\simeq_\phi
\begin{bmatrix}
0.50+0.77i & -0.30-0.08i & 0.0\\
0.09-0.14i & 0.95-0.22i & -0.01i\\
0.0 & 0.0 & -1.11-0.20i\\
\end{bmatrix}
\end{equation*}
where $\mathcal{F}\circ\mathcal{M}(\mathfrak{G})=0.654$. For the case of the Mollifier curve $\mathfrak{M}$ with $\sigma=4.15$, we have
\begin{equation*}
\mathcal{M}(\mathfrak{M})\simeq_\phi
\begin{bmatrix}
0.78+0.62i & -0.07-0.04i & 0.0\\
0.08-0.02i & 0.93-0.35i & 0.0\\
0.0 & 0.0 & -0.10-0.08i\\
\end{bmatrix}
\end{equation*}
with $\mathcal{F}\circ\mathcal{M}(\mathfrak{M})=0.411$. We can thus conclusively state that, with these parameters, the mollifier curve both induces a lesser non-computational transition along with achieving a gate closer to the $CZ$, its $\mathcal{N}$-norm being smaller; this is a demonstration of the possibility that our non-simulative optimization is exactly mappable, in some coherent sense, to a quantum mechanical simulation. In this light, we conclude this work by proposing a question for future consideration.
\\
\\
\textbf{Proposition} Given our definitions for $\mathcal{F}\circ\mathcal{M},\|\cdot\|_\mathcal{N}:\mathcal{N}\to\mathbbm{R}$, let $\mathcal{N}_Q=(\mathcal{N}, \leq_Q)$ and $\mathcal{N}_R=(\mathcal{N},\leq_R)$ be totally ordered sets with $\leq_Q=q^{-1}[\leq_\mathbbm{R}]$ and $\leq_R=r^{-1}[\leq_\mathbbm{R}]$ where $q=(\mathcal{F}\circ\mathcal{M})\times(\mathcal{F}\circ\mathcal{M})$, $r=\|\cdot\|_\mathcal{N}\times\|\cdot\|_\mathcal{N}$ (where we have used the standard Cartesian product of functions), and $\leq_\mathbbm{R}$ is the standard ordering of the real numbers. Then the identity map,
\begin{equation}
\begin{split}
    1_\mathcal{N}:\mathcal{N}_R&\to\mathcal{N}_Q, \\
    \varphi &\mapsto \varphi \label{fut_prop}
\end{split}
\end{equation}
is an order isomorphism.
\\
\\
Here, $\mathcal{N}_Q$ and $\mathcal{N}_R$ represent the set of trajectories ordered by the quantum mechanical fidelity and the norm constructed in this work, respectively. Thus if the identity between them preserves ordering, performing a quantum mechanical simulation to compare trajectories is never necessary, as one can simply apply $\|\cdot\|_\mathcal{N}$.

\section{Conclusion} \label{sec:conclusion}
\subsection{Summary}
In this work, we have addressed the issue of choosing flux trajectories to adiabatically implement a CZ gate in two flux-tunable transmon qubits by introducing a novel method with which one can test and optimize trajectory families. After a brief but rigorous review of this adiabatic implementation, we motivate, through heavy consideration of the non-computational subspace, the definition of a functional norm which places an order relation of diabaticity on the space of possible trajectories, and subsequently apply the technique of constrained variation to arrive at yet another measure for adjustment. After providing examples which justify intuition about the effects of certain trajectories showing the coherence of this model, we simulate implementation on a quantum processor demonstrating success of the non-simulative approach outlined here. 
\subsection{Future Work}
As discussed, there is much room to expand on our methods for optimizing for adiabaticity without simulation. Foremost, different curves may be considered of course, as those presented here are idealized for illustrative purposes; in reality, physical conditions may impose restrictions on flux trajectories, which can be handled by these methods. In addition, the fact that trajectories are optimized entirely mathematically (as opposed to numerically solving the Schrodinger equation or actual physical realization) is also a very helpful tool for design purposes, and can be generalized to different functional norms given different physical focuses. For example, here we placed higher importance in minimizing $\dot{\delta}$ along $\varphi(t)$, which we have mathematically described by raising $\dot{\delta}$ to the second power in Eq. \ref{func_norm}; these powers may be adjusted to influence the importance of physical conditions on solutions. 

Though we have not done so here, it may be possible to construct a norm which gives rise to a \textit{perfect} minimizer: that is, the equation given by constrained variation has a physically realizable and practical solution. Finding a norm with such a solution could make imperfection in adiabaticity, and thus gate error, arbitrarily reducible. Additionally, techniques using variational considerations on functional norms are applicable to a wide variety of cases in the control and design of qubits since time evolution given by the Schrodinger equation often involves trajectoral integration similar to that in this work, e.g. single-qubit pulse design and improvement to the DRAG scheme\cite{superconducting_qubits}.

Lastly, the most important and directly related future consideration is that proposed in Eq. \ref{fut_prop}, which states that it is \textit{entirely} equivalent to use the methods discussed here to optimize flux trajectories as it is to do so simulatively, i.e. the assertion that a trajectory is more adiabatic than another under the $\mathcal{N}$-norm is equivalent to its performing more closely to the ideal $CZ$ under a quantum mechanical simulation. This is a highly advantageous statement, if true, from a purely optimization-based standpoint. In conjunction with a redefinition of the $\mathcal{N}$-norm, in the ideal case these could lead to the determination of the singular best possible trajectory without need for any simulation.

\subsection{Acknowledgements}
We would like to thank V. Narasimhachar for mentorship, discussion, and general research guidance, along with A. Ying for editing, research management, and discussion.

\subsection{Data Availability}
The data and code used to generate figures in this work is available upon request from the authors.

\bibliography{bibliography.bib}


\end{document}